\documentclass[pra,twocolumn,superscriptaddress,showpacs,preprintnumbers,amsmath,amssymb]{revtex4-2}
\usepackage{graphicx}
\usepackage{dcolumn}
\usepackage{bm}
\usepackage{enumerate}
\usepackage{mathrsfs}
\usepackage{amsmath}
\usepackage{epstopdf}
\usepackage[titletoc]{appendix}
\usepackage{hyperref}
\hypersetup{nolinks=false,
urlcolor=cyan, 
citecolor=violet, 
anchorcolor=yellow,
linkcolor=blue, 
citecolor=red,
colorlinks=true}

\begin{document}

\title{Extending Qubit Coherence Time via Hybrid Dynamical Decoupling}
\author{Qi Yao\href{https://orcid.org/0000-0002-0522-2820}{\includegraphics[scale=0.06]{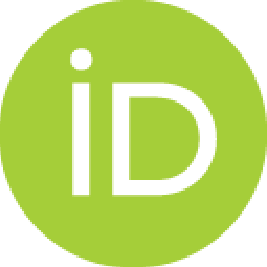}}}
\affiliation{Quantum Science Center of Guangdong-Hong Kong-Macao Greater Bay Area (Guangdong), Shenzhen 518045, China}
\affiliation{Institute of Quantum Precision Measurement, State Key Laboratory of Radio Frequency Heterogeneous Integration, College of Physics and Optoelectronic Engineering, Shenzhen University, Shenzhen 518060, China}

\author{Jun Zhang\href{https://orcid.org/0009-0004-0272-5926}{\includegraphics[scale=0.06]{ORCIDiD.eps}}}
\affiliation{School of Physics and Technology, Wuhan University, Wuhan, Hubei 430072, China}

\author{Wenxian Zhang\href{https://orcid.org/0000-0001-9573-731X}{\includegraphics[scale=0.06]{ORCIDiD.eps}}}
\affiliation{School of Physics, Hangzhou Normal University, Hangzhou, Zhejiang 311121, China}

\author{Chaohong Lee\href{https://orcid.org/0000-0001-9883-5900}{\includegraphics[scale=0.06]{ORCIDiD.eps}}}
\email[Contact author: ]{chleecn@szu.edu.cn}
\affiliation{Institute of Quantum Precision Measurement, State Key Laboratory of Radio Frequency Heterogeneous Integration, College of Physics and Optoelectronic Engineering, Shenzhen University, Shenzhen 518060, China}
\affiliation{Quantum Science Center of Guangdong-Hong Kong-Macao Greater Bay Area (Guangdong), Shenzhen 518045, China}

\date{\today}

\begin{abstract}
Dynamical decoupling (DD) and bath engineering are two parallel techniques employed to mitigate qubit decoherence resulting from their unavoidable coupling to the environment. 
Here, we present a hybrid DD approach that integrates pulsed DD with bath spin polarization to enhance qubit coherence within the central spin model. 
This model, which can be realized using GaAs semiconductor quantum dots or analogous quantum simulators, demonstrates a significant extension of the central spin's coherence time by approximately 2 to 3 orders of magnitude that compared with the free-induced decay time, where the dominant contribution from DD and a moderate improvement from spin-bath polarization.
This study, which integrates uniaxial dynamical decoupling and auxiliary bath-spin engineering, paves the way for prolonging coherence times in various practical quantum systems, including GaAs/AlGaAs, silicon and
Si/SiGe.
And this advancement holds substantial promise for applications in quantum information processing.
\end{abstract}

\maketitle

\section{Introduction.}
Qubits surpass classical bits due to their quantum coherence and entanglement properties, driving the development of various quantum technologies~\cite{Luo2017deterministic,Xu2019Efficient,Guo2021Faster,Yao2011Robust,Violaris2021Transforming,Yao2019Uniaxial}, including cavity dynamic systems~\cite{Duan2001Long,Ma2022Universal,Zhang2022Efficient}, electron~\cite{Loss1998Quantum} and nuclear~\cite{Witzel2007Nuclear,Kurucz2009Qubit,Witzel2007Nuclear} magnetic resonance systems, semiconductor quantum dots (QDs)~\cite{Hanson2007Spins, Kurucz2009Qubit}, cold atomic gases~\cite{Bloch2012Quantum}, and trapped ion systems~\cite{Cirac1995Quantum}. 
However, qubits are inevitably to suffer from different environmental noises~\cite{Glazov2012Spin} (typically $1/f$ noises~\cite{Tosi2017silicon,Hetenyi2019Hyperfine}, phonon-meditated noises~\cite{winkler2003spin}, and white noises~\cite{pla2012single}), leading to a decrease in their effectiveness for quantum computation~\cite{Cirac1995Quantum,Taylor2003Long} and quantum sensing~\cite{Zou2018Beating}. 
Among these platforms, QDs stand out as promising candidates due to their advanced large-scale nano-integration~\cite{Mortemousque2021Enhanced,Borsoi2024Shared}, all-electric control, and ease of coupling with microcavity~\cite{Najer2019Gated}, micromagnet~\cite{Pioro2008Electrically}, electric field~\cite{Khomitsky2020Spin}. 
However, qubits in QDs are easily affected by factors such as acoustic phonon-spin coupling~\cite{Golovach2004Phonon,Semenov2004Phonon}, spin-orbit coupling~\cite{Hideo2002Semiconductor}, hyperfine coupling between qubit and nuclear-spin bath~\cite{Paget1977Low}, and electrode charge noise~\cite{Tosi2017silicon}. 

To mitigate the above factors, various traditional approaches have been proposed~\cite{Hideo2002Semiconductor,Hanson2007Spins,Fischer2008Spin,Reed2016Reduced,chekhovich2017measurement,Bradley2019Ten,paquelet2023reducing}, spanning from the material level to qubit-manipulation techniques.  
Qubits based upon different degrees of freedom have been utilized for various physical applications~\cite{Stano2022Review}, including spin-\cite{Yoneda2018quantum}, charge-\cite{Kim2015Microwave}, and hybrid-qubits\cite{Johnson2005Triplet,Cao2016Tunable}, with fabrication materials such as silicon\cite{Noiri2022Fast,Mills2022Two}, germanium\cite{Hendrickx2021four}, GaAs/AlGaAs\cite{Li2015Conditional}, and silicon metal-oxide semiconductor (SiMOS)\cite{Jones2018Logical}. 
Additionally, promising strategies have been developed for various quantum dots (QDs), including decoherence-free subspaces using charge qubits\cite{Friesen2017decoherence} and hybrid qubits\cite{Farfurnik2021Single}, quantum error correction with three spin qubits\cite{Takeda2022Quantum}, and the quantum Zeno effect\cite{Ahmadiniaz2022Quantum}. 
Ultimately, the hyperfine coupling between spin qubits and nuclear spins remains the primary source of decoherence in spin qubits\cite{Paget1977Low,Johnson2005Triplet}.

To suppress spin-qubit decoherence, particularly that induced by hyperfine coupling in quantum dots (QDs), one naive but costly approach is isotope purification~\cite{Boehme2012Nuclear,Simin2017Locking}. 
Echo-refocusing techniques are employed to decouple the interaction between the central spin and bath spins, including the well-known Hahn echo~\cite{Hahn1950Spin}, Carr-Purcell-Meiboom-Gill (CPMG) and periodic dynamical decoupling (PDD)~\cite{slichter1992principles,Zhang2008Long}, symmetry-based $CDD_\imath$ at the $\imath$-th concatenation level~\cite{Khodjasteh2005Fault,Zhang2008Long}, bias-field-assisted Uni-DD~\cite{Yao2019Uniaxial}, concatenated Uhring DD~\cite{Uhrig2009Concatenated} and quadratic DD~\cite{West2010Near-Optimal} with variable pulse delays~\cite{Uhrig2007keeping}.
However, due to pulse imperfections in experiments, biaxial PDD and $CDD_\imath$~\cite{Khodjasteh2005Fault,Zhang2008Long} lack robustness against amplitude and frequency fluctuations of the driven fields, while Uhrig DD series~\cite{Uhrig2009Concatenated,West2010Near-Optimal,Uhrig2007keeping} are sensitive to pulse timing.
Notably, the Uni-DD and its symmetrized variation~\cite{Yao2019Uniaxial} provide robustness against rotation-angle errors and require fewer pulse overheads.

Parallel to the above storyline, polarization transfer between nuclear spins and electron spins has been realized via optical pumping~\cite{Lampel1968Nuclear,Dobers1988Electrical}, spin-polarized currents in 2DEG~\cite{Dixon1997Dynamic} and dynamical nuclear polarization~\cite{Hogele2012Dynamic,chekhovich2017measurement,chekhovich2020nuclear}. 
These methods can be collectively summarized as the nuclear-spin bath polarization (NsBP) method.
In QDs, the averaged polarization ratio $p$ reaches $\sim$ 0.1-0.7~\cite{Bracker2005Optical,Lai2006Knight,Braun2005Direct,Tartakovskii2007Nuclear,Ono2004Nuclear,Wald1994Local}, effectively creating a clean bath for qubit manipulation or serving as a collective nuclear spin quantum memory.
This naturally raises the question of whether we could combine the NsBP and the Uni-DD proposal to optimize electron-spin-based coherence dynamics.
In this work, we integrate both approaches to enhance qubit coherence using an electron spin in a single QD, where quantum noise is doubly optimized through the hybrid dynamical decoupling method.

\section{The model and the hybrid dynamical decoupling method}
\begin{figure}[b]
\centering
\includegraphics[width=3.0in]{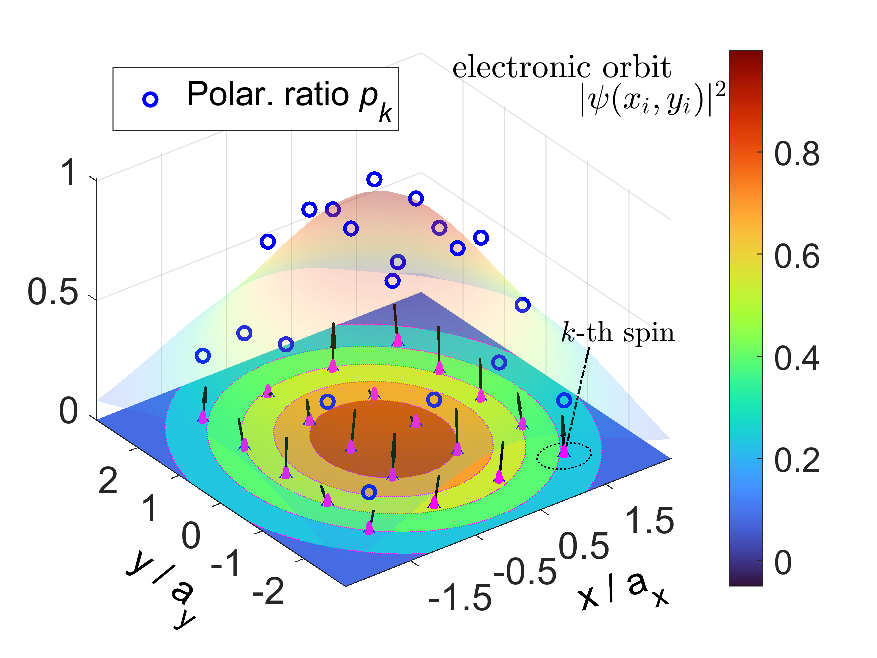}
\caption{\label{fig:ra} Central spin model and bath spin polarization. One of realizing central spin model is semiconductor QD structure, an electron spin $\mathbf{S}$ interacting with surrounding $N=20$ nuclear spins $\mathbf{I}$ in a QD at a bias magnetic field $\mathbf{b}_e$, the electron is confined in a Gaussian-like form. The electronic spatial orbital is represented by its shaded surface, with the electronic density $|\psi(x_i,y_i)|^2$ depicted as a contour envelope in normalized 2D spatial coordinates. The nuclear spins are also sketched within the $x$-$y$ contour envelope. When partially polarizing the nuclear-spin bath~\cite{Hogele2012Dynamic,chekhovich2017measurement,chekhovich2020nuclear} (with averaged polarization ratio $p=0.60$), the polarization ratio $p_k$ (blue circle) of the $k$-th spin is observed through its $z$-component expectation $2\langle I_k^z \rangle$.}
\end{figure}

\textit{Central spin model.} \textbf{--}
We consider a central spin $\textbf{S}$ that coupled to the external magnetic field $\textbf{b}_{e}$ and interacts with bath spins $\textbf{I}_k$, 
\begin{align}\label{eq:e2}
H_0=\textrm{g}_e^*\mu_B\textbf{b}_{e}\cdot \textbf{S}+\sum_{k=1}^NA_k\textbf{S}\cdot\textbf{I}_k,
\end{align}
where the first term is Zeeman energy with the effective Land\'{e} factor $\textrm{g}_e^*$ and the Bohr magneton $\mu_B$, the second term is the hyperfine interaction between the central spin and $N$ bath spins with the coupling strength $A_k$ at the $k$-th bath spin. 
This model can be realized by analogy (such as gate-defined GaAs semiconductor QD system~\cite{de2003Theory,Witzel2006Quantum,Coish2004Hyperfine}, superconducting circuit~\cite{Chowdhury2024Enhancing}, Rydberg atom~\cite{Dobrzyniecki2023Quantum}), and digital quantum simulation~\cite{Heras2014Digital,Salathe2015Digital}.
The dipolar-dipolar interaction between bath spins is weak and can be neglected in many individual QDs~\cite{Paget1977Low,yao2024note}, see the Appendix~\ref{sup1}. 
In our numerical simulation, bath spins form a 2D $4\times5$ spin lattice with $A_k\varpropto\exp[-(x-x_0)^2/\omega_x^2-(y-y_0)^2/\omega_y^2]$ in a 2D Gaussian-like form (as illustrated in Fig.~\ref{fig:ra}). 
The widths are $\omega_x$ and $\omega_y$, and the mismatched shifts are $x_0=0.1a_x$ and $y_0=0.1a_y$~\cite{Dobrovitski2006Long,Zhang2008Long,Yao2019Uniaxial}. 
Two distributions of $A_k$ are considered: (i) a normal width of $A_k$ that changes between 0.309 and 0.960 with $\omega_x/a_x=3/2$ and $\omega_y/a_y=2$ along the $x$ and $y$ axis, and (ii) a narrow width of $A_k$ varies from 0.096 to 0.922 with $\omega_x/a_x=3/(2\sqrt2)$ and $\omega_y/a_y=\sqrt2$. 
These two distributions can be achieved experimentally by controlling the QD size~\cite{Murray1993Synthesis} or using digital quantum simulation~\cite{Salathe2015Digital}.

In our scheme, the (central) electron spin and (bath) nuclear spins are initially prepared in a product state with a sudden approximation~\cite{Sakurai1995Modern}:
$|\psi(0)\rangle=|\psi_e(0)\rangle\otimes|\psi_b(0)\rangle$ with $|\psi_e(0)\rangle$ and $\psi_b(0)\rangle$ respectively denoting the electron-spin and nuclear-spin states. 
The electron spin is usually prepared in longitudinal states: spin up $\left|\uparrow\right\rangle$ and spin down $\left|\downarrow\right\rangle$, or transverse states: $\left|x\right\rangle=\frac{1}{\sqrt{2}}(\left|\uparrow\right\rangle+\left|\downarrow\right\rangle)$ and $\left|y\right\rangle=\frac{1}{\sqrt{2}}(\left|\uparrow\right\rangle+\textit{i}\left|\downarrow\right\rangle)$. 
These states involve relaxation and dephasing processes and can be exploited to benchmark the decoupling efforts of different DD sequences~\cite{Hahn1950Spin,Uhrig2007keeping,Khodjasteh2005Fault}. 
The collective spin bath state has the form of $|{\psi_b(0)}\rangle \langle \psi_b(0)|=\prod_{k=1}^N (\mathbb{I}_2/2+p_kI_k^z)$, where $\mathbb{I}_2$ is the identity operator and $p_k\equiv 2\langle I_k^z \rangle$ is the polarization ratio of the $k$-th bath spin with an observed value $\langle I_k^z\rangle$. 
When the spin bath is in thermal equilibrium, its averaged polarization ratio $\textit{p}=\sum_{k=1}^N p_k/N$ equals 0, and $p_k$ is independently and identically distributed with values ranging from -1 to 1, with the norm $\| |\psi_b(0)\rangle \|= 1$.

When collective (nuclear) spin bath states are used as storage entities~\cite{Taylor2003Long,Denning2019Collective}, their storage limit (coherence time) can be increased by polarization enhancement~\cite{Ding2014High,Jing2023Noise}. To engineer a nuclear spin state, both static~\cite{slichter1992principles,Coish2004Hyperfine,Deng2006Analytical} and dynamical~\cite{Reilly2008suppressing,Xu2009Optically,Flisinski2010Optically} nuclear polarization (i.e. DNP) have been studied theoretically and experimentally. 
Here, we adopt a three-step method under the independent spin approximation~\cite{Petta2008Dynamic,Zhang2010Protection,Wu2016Inhomogeneous}.  
The nuclear-spin bath can be prepared as inhomogeneously polarized state~\cite{Taylor2003Controlling,Dobrovitski2006Long,Ding2014High,Wu2016Inhomogeneous}, $| \psi_b^{'}(0)\rangle=c\exp(\beta\sum_kA_k^2I_k^z) |\psi_b(0)\rangle$,
where the prefactor $c$ is the normalized constant, and the polarized parameter $\beta$ is mainly related to the mixing time (see the DNP details in Appendix~\ref{sup1}). 
The distribution of $p_k$ [Fig.~\ref{fig:ra}] depends on its hyperfine strength at the $k$-th nuclear spin in the QD.

To monitor the dynamics of a central spin state, $\rho_e(0)=|\psi_e(0)\rangle \langle \psi_e(0)|$, we benchmark its fidelity,
\begin{eqnarray}\label{eq:e5}
F=\operatorname{Tr}\left\{\rho_e(0) \operatorname{Tr}_n[\rho(t)]\right\}.
\end{eqnarray}
We need to single out the worst case among the four initial states $\{\left|\uparrow\right\rangle, \left|\downarrow\right\rangle, \left|x\right\rangle, \left|y\right\rangle\}$, $F_w=\min _{\rho_e(0)}(F)$. 
Note that $|\rho(t)\rangle=\exp(-iHt)\rho(0)\exp(iHt)$, $\operatorname{Tr}_n[\rho(t)]$ denotes the trace of the spin bath, and $\operatorname{Tr}\{\rho_e(0) \cdot \cdot\cdot\}$ means that the initial state $\rho_e(0)$ is used as the measure operator.
Here, we employ the Chebyshev polynomial method~\cite{Dobrovitski2003Efficient} to solve the long-term quantum dynamics~\cite{Zhang2006Hyperfine,Zhang2008Long,Yao2019Uniaxial}.


\begin{figure}[!htpb]
 \centering
\includegraphics[width=3.35in]{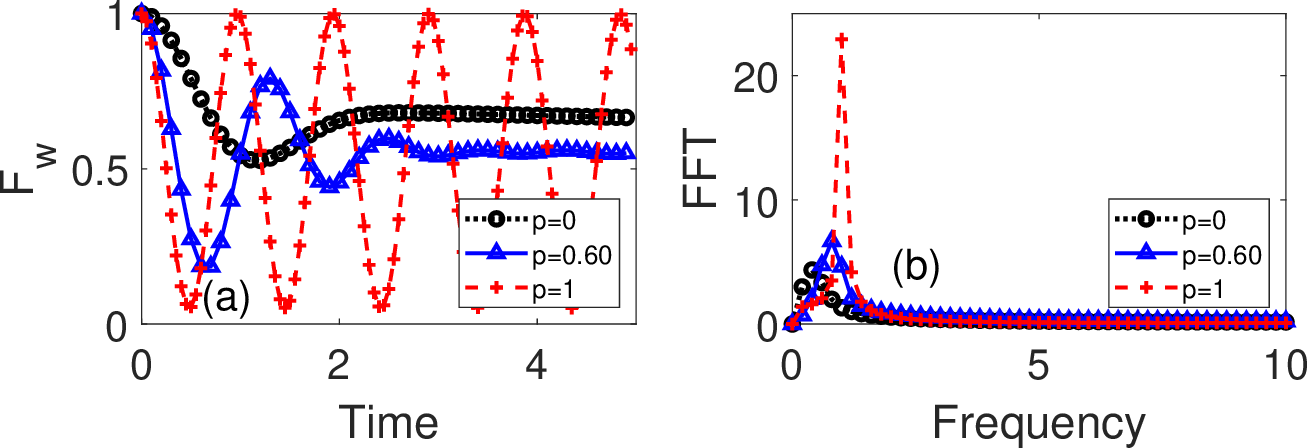}
\caption{\label{fig:rb}  The dynamic of the electronic spin under various polarized nuclear-spin bath state without the external magnetic field $\textbf{b}_{e}$, where time is measured in units of inverse $A_k$ [cf. Ref.~\cite{Coish2004Hyperfine}]. (a) The minimal fidelity $F_w$ as a function of three polarization cases: the no polarized nuclear-spin bath with averaged polarization ratio $p\simeq$0 (black circle line), the partial polarized nuclear-spin bath with $p$=0.60 (blue triangle line), the full polarized nuclear-spin bath with $p$=1 (red plus line) vs time $t$, respectively. (b) The corresponding Fourier transform of the data from the left panel.}
\end{figure}

\emph{Hybrid DD protocol.} \textbf{--}
The Uni-DD protocol~\cite{Yao2019Uniaxial} consists of a relatively strong bias field $\mathbf{b}_e^z$ along $z$-axis and a sequence of instantaneous $\pi$ pulses ``Y" along $y$-axis.
And the pulse delay satisfies the magic condition $\tau=2\pi/\omega$, where $\omega=\gamma b_e^z$~\cite{Yao2019Uniaxial}.
The hybrid DD protocol differs from the Uni-DD protocol~\cite{Yao2019Uniaxial}.
Both Uni-DD and hybrid-DD can be briefly described as follows,
\begin{equation}\label{eq:e1}
\begin{aligned}
& \text { Uni-DD }: \left[Y U_\tau Y U_\tau\right]^L \text { based on }\left|\psi_b(0)\right\rangle, \\
& \text { Hybrid DD }:\left[Y U_\tau^{\prime} Y U_\tau^{\prime}\right]^L \text { based on }\left|\psi_b^{\prime}(0)\right\rangle,
\end{aligned}
\end{equation}
where $U_\tau$ ($U_\tau^{\prime}$) denotes the dynamical operator of Eq.~\ref{eq:e2} and $L$ is the cycle number in total dynamical evolution, and its evolution time is $2L\tau$. The polarized nuclear-spin bath state $|\psi_b^{'}\rangle$ is obtained by the NsBP method.

For the (central) electron spin in a QD, its free coherent dynamic decays rapidly, primarily due to the fluctuating Overhauser field or nuclear spin noise.  
We first consider the impact of the polarized nuclear-spin bath on electronic spin dynamic in a QD~\cite{Bluhm2010Enhancing}. 
As shown in Fig.~\ref{fig:rb}(a), the electron spin decoheres due to the raw Overhauser field~\cite{Coish2004Hyperfine,Yao2006Theory,Glazov2012Spin} (i.e. the averaged polarization ratio $p\simeq0$), causing the fidelity to drop under quasistatic approximation~\cite{Merkulov2002Electron}.
With the collective nuclear-spin state $|\psi_b(0)\rangle$ being gradually polarized, the evolution of electron spin is dominated by the equivalently (static) bias field of non-zero $\overline{h_o^{z}}=\sum_k A_k p_k/2$, along with the smaller fluctuation scope $\Delta h_o^{z}=\sqrt{\sum_k A_k^2(1-p_k^2)}/2$ at the polarization direction~\cite{vink2009locking,Laird2007Hyperfine,Tenberg2015Narrowing}, which is proportional to the width of the half-height of the peak [see Fig.~\ref{fig:rb}(b)]. 
Thus, the longitudinal states only undergo slight relaxation, and the transverse states undergo an oscillatory decay. 
As $p$ increases, the transverse states oscillate intensively with the approximated Larmor angular frequency $\omega_z=\textrm{g}_e^*\mu_B\overline{h_{o}^z}/\hbar$.
Until the polarization ratio $p$ saturates at 1, the cycle number of the Larmor process increases significantly [Fig.~\ref{fig:rb}(a)], and these worst states are still dynamically preserved. 

\begin{figure}[!htpb]
 \centering
\includegraphics[width=3.35in]{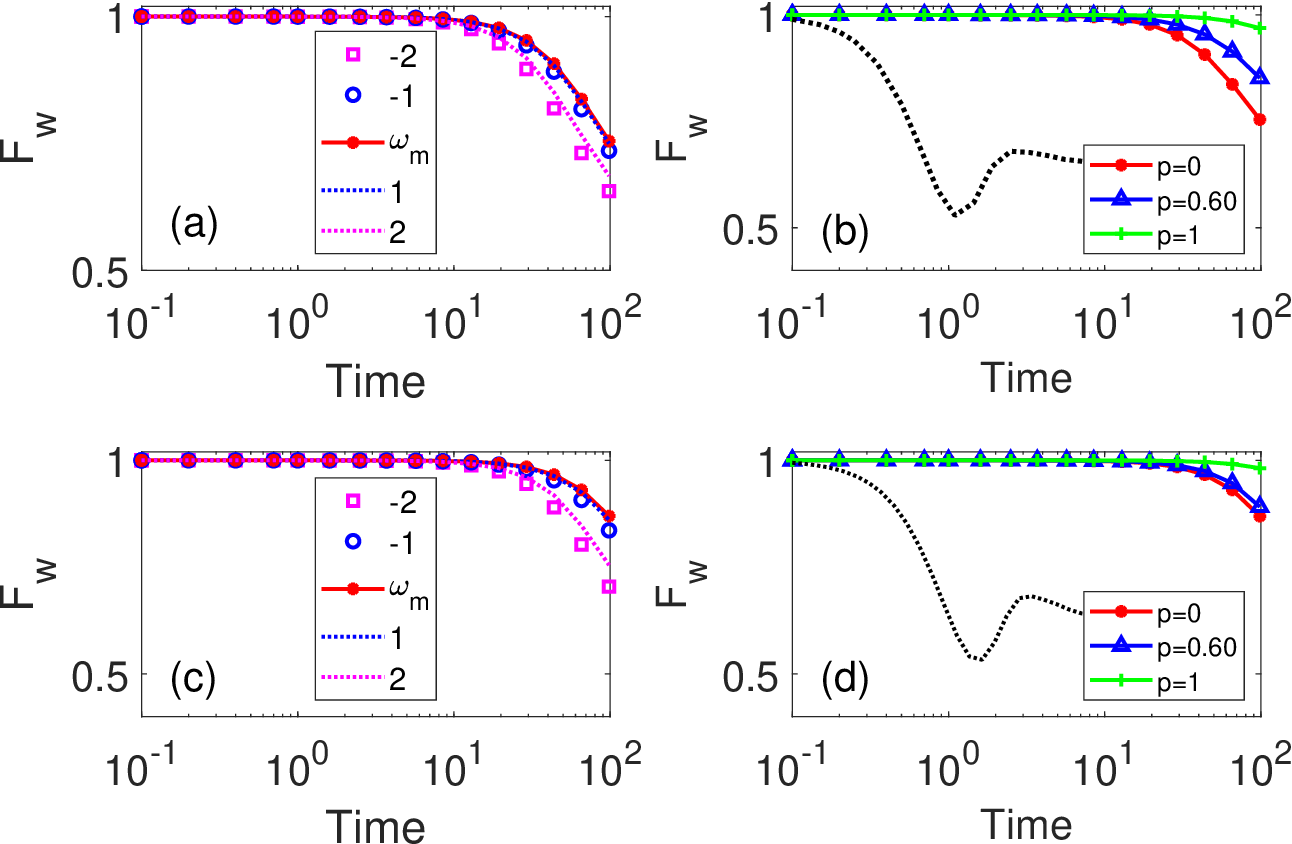}
\caption{\label{fig:rc} The effects of the Uni-DD and the hybrid DD protocols  is analyzed by narrowing the size of the QD or polarizing its nuclear-spin bath. The minimal fidelity $F_w$ as a function of external magnetic field $(\omega-\omega_m)/\gamma$ or in three polarized nuclear-spin bath cases.  For the Uni-DD protocol, two types of $A_k$-distribution are considered: the normal case (a) and the narrow case (c). The magic condition is satisfied with $\omega_m=2\pi/\tau$ (red solid line) or not satisfied at $\omega-\omega_m=\pm2$ (magenta square or dashed line), $\pm1$ (blue circle or dashed line). For the hybrid DD protocol, the normal (b) and narrow cases (d) with the same $\tau$ are illustrated in three scenarios: the unpolarized nuclear-spin bath with bath polarization ratio $p$=0 (red star line), the partially polarized nuclear-spin bath with $p$=0.60 (blue triangle line), and the fully-polarized nuclear-spin bath with $p$=1 (green plus line). The pure free-evolution (black dashed line) is added for easier comparisons in the log-linear scale, where the pulse delay $\tau=0.05$. The timestamp is identical to that in Fig.~\ref{fig:rb}.}
\end{figure}

It is worth mentioning that the above process, combined with the slowly varying nuclear-spin bath~\cite{yao2024note} in QDs, can be reversed to use nuclear spins for storing electron spin state. 
Based on collected and polarized nuclear spins, there appears a quantum memory scheme~\cite{Taylor2003Long}, in which the ``read-write" mode between the electron spin and the collected nuclear spins is accomplished by roughly half Rabi oscillation cycles. 
Its storage performance can be enhanced~\cite{Ding2014High,Denning2019Collective,Jing2023Noise} through NsBP~\cite{Hogele2012Dynamic,chekhovich2017measurement,chekhovich2020nuclear} method and isotope purification~\cite{Boehme2012Nuclear,Simin2017Locking,Bourassa2020Entanglement}.
Its storage fidelity is inevitably limited by fluctuations in the Overhauser field. 
Moreover, the ``read-write" process that uses the interaction between electron spin and nuclear spin requires intricate control techniques (see the Hamiltonian engineering sequence, $[XXYY]^n$ in~Ref.\cite{Jing2023Noise}), and achieving a high bath polarization ratio is indeed challenging. 
In the Appendix~\ref{sup3}, we present a quantum storage scheme based on electron spin [Fig. S1] using the hybrid DD method. 

Analysis of $\Delta h_o^{z}$ reveals that, beyond narrowing the Overhauser field distribution~\cite{vink2009locking,Laird2007Hyperfine,Tenberg2015Narrowing}, other methods such as engineering sample size of nuclear-spin bath are experimentally feasible~\cite{Murray1993Synthesis}. 
Supposing the collective nuclear spins are not polarized using the NsBP method, reducing the QD size concentrates the distribution of $A_k$, thereby reducing the fluctuation range of the Overhauser field, $\Delta h_o^{x,y,z}=\sqrt{\sum_k A_k^2}/2$. 
For example, when the system evolves at time $t=100$ in the Uni-DD sequence, the minimal fidelity $F_w$ in the narrow case is slightly higher than in the normal cases at $\omega-\omega_m=0$ [Fig~\ref{fig:rc}(a, c)]. 
Hence, one can infer that the coherence time of the electron spin is extended longer with the Uni-DD in the central spin model when the QD size is more narrowly confined.
Note that, compared to the magnetic dipolar-dipolar interaction between nuclear spins included in the central spin model~\cite{Yao2019Uniaxial}, the value of $F_w$ in this work decreases slightly. 
This reduction is attributed to the omission of the correlation effects of the nuclear-spin bath~\cite{Paget1977Low,yao2024note}.

We emphasize two points about the above two methods: the NsBP method or narrowing the size of QD. 
Firstly, the fluctuation scope of $\mathbf{h}_{0}$ can be efficiently reduced by engineering its bath spins. 
However, the high nuclear polarization brings some difficulties in experiments~\cite{Lampel1968Nuclear,Dobers1988Electrical,Dixon1997Dynamic}. 
Secondly, although pulsed biaxial DD sequences such as biaxial $XY8$~\cite{Ryan2010Robust}, $CDD_\iota$~\cite{Zhang2006Hyperfine}, concentrated Uhring DD\cite{Uhrig2009Concatenated}, and quadratic DD\cite{West2010Near-Optimal} can decouple $\mathbf{h}_{o}$ to some extent in a nested way~\cite{Zhang2006Hyperfine,Ryan2010Robust,Souza2011Robust}, they need to avoid non-orthogonal axial directions, rotation angle errors, and imprecise timing.

In the central spin model, the decoupling effort and robustness of Uni-DD can be clearly observed in both the normal and narrow cases [Fig~\ref{fig:rc}(a, c)].
Compared to the pure free-evolution time, the coherence time of electron spin could be extended by approaching two orders of magnitude.
When the magic condition does not match, as the deviation of $\omega-\omega_m$ increases, where $\omega-\omega_m=\pm1$ and $\omega-\omega_m=\pm2$, the worst fidelity $F_w$ falls rapidly in the same evolution time. 
When the nuclear-spin bath is polarized, as the polarization ratio $p$ increases, the overlap between the red and green lines in Fig.~\ref{fig:rc}(b, d), the desired effect is achieved across various sizes of QDs, compared to the two cases with bare Uni-DD protocol~\cite{Yao2019Uniaxial}. 
Therefore, our hybrid DD method effectively combines these two approaches to significantly enhance the coherence time of the electron spin.

We further qualitatively analyze the decoupling effort of our hybrid DD method. 
In one period $U_{2\tau}$, $U_{2\tau}\approx\exp \{-i2\tau[S_y(\overline{h_o^z} h_o^y)/\omega_m^{\prime}+i(h_o^x h_o^y-h_o^yh_o^x)/(4\omega_m^{\prime})]\}$ in our hybrid DD protocol performs better than $U_{2\tau}\approx\exp \{-i2\tau[S_y(h_o^z h_o^y+h_o^yh_o^z)/(2\gamma b_e^z)+i(h_o^x h_o^y-h_o^yh_o^x)/(4\gamma b_e^z)]\}$ in the Uni-DD protocol~\cite{Yao2019Uniaxial}, where $\overline{h_o^z}$ is the narrowed Overhauser field with the polarized nuclear-spin bath
state $|\psi_b^{'}\rangle$, and the effective Rabi frequency $\omega_m^{\prime}=\gamma(b_e^z+b_o^z)$.
The derivation in Bloch rotation theory can be found in the Appendix~\ref{sup2}.
This indicates that the hybrid DD protocol outperforms the Uni-DD protocol and their ability to suppress the Overhauser field $\mathbf{h}_o$ is unaffected by the initial state of the electron spin.

\begin{figure}[!htpb]
 \centering
\includegraphics[width=3.0in]{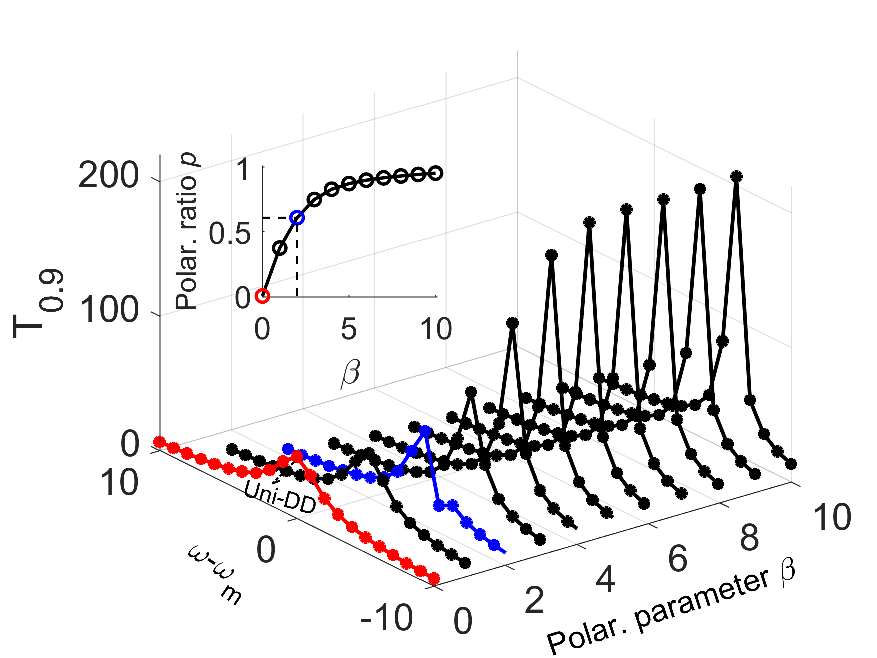}
\caption{\label{fig:rd} In the hybrid DD protocol, the characteristic coherence time $T_{0.9}$ where the minimal fidelity $F_w$ drops to 0.9, is benchmarked as the function of $\omega-\omega_m$ and the polarized parameter $\beta$ in the normal case. In the vicinity of $\omega_m\approx 126$, scanning $\gamma b_e^z$ at intervals of 1 to capture the Overhauser field $\overline{h_o^z}$, thereby matching the magic condition in different polarization situations. The inset gives the nuclear-spin bath polarization ratio $p$ vs the polarized parameter $\beta$.}
\end{figure}

The advantage of our hybrid DD scheme lies in the modified magic condition. 
Specifically, the equivalent magnetic field $\textbf{b}_o^z$ along $z$-axis partially offsets the bias field $b_e^z$, effectively shifting the magic condition of the Uni-DD scheme. 
Consequently, the magic condition becomes $b_e^z+b_o^z=\omega_m^{\prime}/\gamma$, indicating that enhanced decoupling performance can be achieved with a reduced magnetic field strength. 
For example, considering the case of a normal distribution of $A_k$, one can still compensate for the deviation of the external bias field $\gamma b_e^z-\omega_m$ in the original ``Uni-DD" proposal by using a fixed pulse delay $\tau$ and $\omega_m=2\pi/\tau$. 
In Fig.~\ref{fig:rd}, as the nuclear-spin bath is gradually polarized by increasing $\beta$, the peak of $T_{0.9}$ increases significantly when the magic condition is matched, otherwise $T_{0.9}$ drops. 
At the same time, its peak position also moves and decreases as the bath polarization ratio $p$ increases, with a narrower peak width due to the smaller scope of $h_o^{z}$~\cite{Dobers1988Electrical}. 
A similar behavior has been confirmed in Ref.~\cite{Coish2004Hyperfine}. 
It is worth noting that, compared to the free evolution time, the coherence time of the central spin can be extended significantly by approximately 2 to 3 orders of magnitude.

\textit{Experimental feasibility.} \textbf{--}
In many solid-state systems, such as GaAs/AlGaAs~\cite{Petta2005Coherent,shulman2012demonstration,Malinowski2017notch}, silicon and Si/SiGe~\cite{Noiri2022Fast,Mills2022Two}, and germanium~\cite{Hendrickx2021four}, the hyperfine coupling between electron (hole) spin and nuclear spins typically dominates over the magnetic dipole-dipole interactions between nuclear spins.  
Although the specific coupling forms vary between these QD nanostructures, and the bath spin number ranges from $N\sim 10^4-10^6$, techniques such as dynamical decoupling (DD) or dynamic nuclear polarization (DNP)~\cite{Petta2008Dynamic} are broadly applicable to mitigate the distortive effects of nuclear spins. 
Furthermore, advanced methods such as Uni-DD and its combination with the NsBP technique can be effectively utilized in these systems. 
Or they can be demonstrated in a digital superconducting quantum circuit simulator.

\section{conclusions}
In summary, we propose a hybrid DD protocol that significantly extends qubit coherence by combining the Uni-DD and NsBP techniques. 
Our numerical results demonstrate that this combination further extends electron spin coherence time compared to using either the Uni-DD or the NsBP method alone. 
Physically, the Overhauser field generated by the NsBP method shifts the magic condition of the Uni-DD scheme. 
By scanning fixed pulse intervals, we identified the magic condition of the renormalized external bias field, it can also be utilized to measure the Overhauser field. 
Our scheme is applicable to other central spin systems, such as silicon, doped-ion crystals~\cite{Beckert2024Emergence}, GaAs/AlGaAs~\cite{Petta2005Coherent}, even in analog quantum simulators.

\emph{Acknowledgements.} \textbf{--}
This work is supported by the National Key Research and Development Program of China (Grant No. 2022YFA1404104), the National Natural Science Foundation of China (grants No. 12025509, No. 12274331, No. 91836101, No. 92476201, and No. 12404272), the NSAF (grant No. U1930201), and the Guangdong Provincial Quantum Science Strategic Initiative (GDZX2305006, GDZX2304007, GDZX2405002).

\begin{appendix}
\vspace{10 pt}

\section{Quantum dot model and the nuclear-spin polarization method}\label{sup1}
For a gate-defined GaAs semiconductor qunautm dot (QD) system~\cite{de2003Theory,Witzel2006Quantum,Coish2004Hyperfine,Wang2012Self,Merkulov2002Electron}, usually placing an external magnetic field $\textbf{b}_{e}$ to couple it, and its Hamiltonian is composed of four parts: $H=H_{ez}+H_{nz}+H_{en}+H_{nn}$, which includes the central (electron) spin Zeeman energy $H_{ez}=\textrm{g}_e^*u_B\textbf{b}_{e}\cdot \bf{S}$, the Zeeman energies of nuclear-spin bath $H_{nz}=\textrm{g}_nu_n\textbf{b}_{e}\sum_{k} \textbf{I}_k$, the hyperfine coupling between between electron spin and nuclear-spin bath $H_{en}=\sum_{k} (8\pi/3)\textrm{g}_e^*\mu_B\textrm{g}_n\mu_n | \psi(\textbf{x}_k) |^2 \textbf{S}\cdot\textbf{I}_k$ with the strength $A_k=(8\pi/3)\textrm{g}_e^*\mu_B\textrm{g}_n\mu_n | \psi(\textbf{x}_k) |^2$, and the magnetic dipole interactions between intra nuclear-spins $H_{nn}=\sum_{k} \sum_{\ell<k} \Gamma_{k \ell} (\mathbf{I}_k \cdot \mathbf{I}_{\ell}-3 I_k^z I_{\ell}^z)$ with the coupling strength $\Gamma_{k \ell}$. Here, $\bf{S}$ is electron spin operator and $\bf{I}_k$ is k-th nuclear spin operator.
$\textrm{g}_e^*$ and $\textrm{g}_n$ are the effective electron and nucleon $g$\textbf{-}factor, respectively. $u_B$ is the Bohr magneton, the electronic density $\psi(\textbf{x}_k) |^2$ at nuclear spin site $\textbf{x}_k$ determines the hyperfine coupling strength between electron spin and the $k$\textbf{-}th nuclear spin.

In GaAs QD with $10^4$ nuclei, $A_k\approx 90 \mu$eV~\cite{Taylor2003Long}, $g_e^* \simeq -0.44$, and  $g_I$ depends on isotype nuclear spins [$g_I(^{69}Ga)=1.344$, $g_I(^{71}Ga)=1.708$, $g_I(^{75}As)=0.96$]. Generally, $H_{nz}$ could be neglected due to the significant difference in the effective mass of electron and nuclear mass.
After unitizing the parameters of $\textrm{g}_e^*$, $\textrm{g}_n^*$, $u_B$, and $u_n$,
the above system is modeled by
\begin{equation}\label{eq:s1}
H =\textrm{g}_e^*\mu_B\textbf{b}_{e}\cdot \textbf{S} + \mathbf{S} \cdot \sum_{k=1}^{N} A_k \mathbf{I}_k + \sum_{k=1}^{N} \sum_{j \neq k} \Gamma_{jk} (\mathbf{I}_k \cdot I_j - 3\mathbf{I}_{kz} \mathbf{I}_{jz}).
\end{equation}
Here, the hyperfine coupling constant \(A_k\) is estimated at $\geq 0.1$ neV~\cite{Taylor2007Relaxation} , and the dipolar magnetic coupling strength \(\Gamma_{jk}\) $\sim 0.01$neV~\cite{Taylor2007Relaxation} between nuclear spins. The third term of Eq.~\ref{eq:s1} results in the slow dynamic of nuclear-spin bath. Note that in the Uni-DD method~\cite{Yao2019Uniaxial}, the effect of this term has been considered in the long-time dynamics, comparing the result of Fig.~3(a) in main text and Fig.~3(a) in Ref.~\cite{Yao2019Uniaxial} and can be neglected. Hence, the effective model becomes Eq.~1 in the main text , which also is called the central spin model~\cite{slichter1992principles,Hackmann2014Spin,Frohling2018Nuclear}.

For Eq.~\ref{eq:s1} without the third term, its intrinsic timescale is estimated $T^*\simeq 1/\sqrt{\sum_k  4A_k^2I_k(I_k+1)/3}$~\cite{Frohling2018Nuclear}, where $I_k$ is quantum number of nuclear spin $\bf{I}$. In addition, the decoherence of electron spin in QD system is closely related to the Overhauser field and its fluctuation scope, both can be defined as, $\mathbf{h}_o^\alpha=\sum_k A_k \mathbf{I}_k^\alpha$, $\Delta h_o^{\alpha}=\sqrt{\langle h_o^\alpha h_o^\alpha \rangle - \langle \mathbf{h}_o^\alpha \rangle^2}$,  where $\alpha=x, y, z$, and $h_o^\alpha h_o^\alpha=\sum_j \sum_k A_j A_k \mathbf{I}_j^\alpha \mathbf{I}_k^\alpha$. For a nuclear spin bath with quantum number of $I=1/2$, and two useful formulas are introduced, $[I^x, I^y] = iI^z$, $\left\langle\left(I_k^\alpha\right)^2\right\rangle=\frac{1}{3}\left\langle I_k^2\right\rangle=\frac{1}{3} \cdot \frac{3}{4}=\frac{1}{4}$. When the nuclear-spin bath is in thermal equilibrium, the in-plane fluctuation scope can be derived,
\begin{eqnarray}\label{eq:s2}
\begin{aligned}
\Delta h_o^x & =\Delta h_o^y=\frac{1}{2} \sqrt{\sum_k A_k^2}.
\end{aligned}
\end{eqnarray}
When the nuclear-spin bath is partially polarized along $z$-axes (i.e., $\langle\mathbf{I}_k^{z}\rangle =p_k/2$), the out-of-plane fluctuation scope has
\begin{eqnarray}\label{eq:s3}
\Delta h_o^z & =\frac{1}{2} \sqrt{\sum_k A_k^2\left(1-p_k^2\right)},
\end{eqnarray}
the $\Delta h_o^x$ and $\Delta h_o^y$ keep unchanged.

\textit{The static NP method.}\textbf{--}
In nuclear magnetic resonance (NMR) experiments or magnetic resonance imaging (MRI) techniques~\cite{Abraham1973Nuclear,haeberlen1976high,slichter1992principles}, the resonance position and width of the signal peak, are directly affected by the polarization degree of nuclear spins. At the same time, since the Zeeman energy of nuclear spin in an external magnetic field is much smaller than that of an electron spin ($u_B/u_n \sim 10^3$), the polarization of nuclear spins is generally more difficult to carry out than that of the electron spins. For a nuclear spin ensemble at temperature $T$ with $H_{nz}=g_n \mu_n b_e^z \sum_k I_k^z$, the magnetization is given by $\left\langle M_z\right\rangle \approx \frac{g_n^2 \mu_n^2 I(I+1)}{3 k T} b_e^z$.
This formula indicates that achieving high polarization of nuclear spins requires both a stronger external magnetic field and a much lower temperature. For example, for $^{69}Ga$ and $^{75}As$ nuclei in an external magnetic field of $b_e^z=10$ T and at $T=20$ mK, the polarizability is $\langle I_z\rangle=0.27$ and 0.14, respectively. Each isotope spin's static polarization ratio is the same, hence it is referred to as homogeneous polarization in static nuclear polarization.

However, the situation is different in QD systems due to the strong hyperfine coupling between electron spins and nuclear spins. In these systems, nuclear spin polarization is achieved through methods such as optical pumping~\cite{Bracker2005Optical}, spin-flip Raman scattering~\cite{Debus2014Spin1,Debus2014Spin2,Debus2015Nuclear}, and the quantum Hall effect~\cite{Dixon1997Dynamic}. These methods are collectively referred to as dynamic nuclear polarization (DNP). In the following, we outline the basic principles of DNP, which can be found in more detail in Refs.~\cite{Koppens2005Control,Ding2014High,Wu2016Inhomogeneous,Yao2010Feedback}.

\textit{The DNP method.}\textbf{--}
The DNP process involves polarization transfer between electron spins and nuclear spins, consisting of repeated cycles, each comprising three main steps:
\begin{enumerate}[(1)]
  \item Initialization: A polarized electron spin, $\rho(0)=|\uparrow\rangle\langle\uparrow|$, is injected into the QD. Considering a single nuclear spin, $\rho_n(0)=(\mathbb{I}_2/2+p I^z)$ with the unknown polarized ratio $p$, the total system's density matrix is $\rho(0)=\rho_e(0) \otimes \rho_n(0)$.
  \item Evolution: The total system evolves to the final state after a specified mixing time $\tau_m$, $\rho_e(\tau_m)=\exp(-iH\tau_m)\rho_e(0) \exp(iH\tau_m)$. This state can be decomposed into: $\rho(\tau_m)  =|\uparrow\rangle_e\langle\uparrow| \otimes \rho_n^{\uparrow \uparrow}+| \uparrow \rangle_e \langle\downarrow| \otimes \rho_n^{\uparrow \downarrow} +|\downarrow\rangle_e \langle\uparrow |\otimes \rho_n^{\downarrow \uparrow}+| \downarrow \rangle_e\langle\downarrow| \otimes \rho_n^{\downarrow \downarrow}$. During this process, the electron spin polarization is transferred to the nuclear spins through hyperfine coupling (the dipolar nuclear interaction is sufficiently weak and can be neglected), allowing the nuclear spins to gain polarization.
  \item Ejection: The electron is ejected from the quantum dot, i.e., tracing out the electron spin,  $\rho_n(\tau_m)=\langle\uparrow|\rho(\tau_m)|\uparrow\rangle+\langle\downarrow|\rho(\tau_m)|\downarrow\rangle=\rho_n^{\uparrow \uparrow}+\rho_n^{\downarrow \downarrow}$, leaving the nuclear spins more polarized than before.
\end{enumerate}
The above process can be solved under the independent spin approximation (ISA). After the mixing time $\tau_m$, the polarization of the electron spin becomes $p_s(t)=(1 / 2)[(1+p)+(1-p)\cos(At)$. The corresponding change in the nuclear spin polarization is $\Delta p(\tau_m)\equiv 1-p_s(t)=(1-p)[(1-\cos(At)]/2$. Besides the mixing time $\tau_m$, the total cycle time $\tau_{tol}$ may include injection time, ejection time. If $\tau_{tol}$ is enough short to ensure that the gained polarization of the nuclear spin $\Delta p \ll1$,  the above difference equation can be approximated as the differential equation $\frac{dp}{dn}\approx\frac14A^2\tau_{tol}^2(1-p)$. The nuclear polarization \( p(t = n\tau_{tol}) \) after \(n\) cycles of DNP can be expressed as:
\begin{equation}\label{eq:s4}
p(t = n\tau_{tol}) = 1 - e^{- \frac{A^2 \tau_{tol}}{4} t}.
\end{equation}
This equation shows that nuclear polarization grows exponentially with the number of DNP cycles and is proportional to the square of the hyperfine coupling constant. Defining $\beta=\tau^2n/4$, the polarization ratio of the nuclear spin $p \approx \beta A^2$ when $\beta A^2\ll1$, and $p\approx\tanh(\beta A^2/2)$ when $\beta A^2\gg 1$, the validity of the ISA is verified in Ref.~\cite{Wu2016Inhomogeneous}.
When many nuclear spins are involved in the DNP process, numerical simulations within the core-skirt model~\cite{Zhang2010Protection} reveal that nuclear polarization is highly inhomogeneous ($p_k \propto \beta A_k^2$). Consequently, strongly coupled nuclear spins polarize much faster than weakly coupled ones, as determined by the $\psi(\textbf{x}_k)$-dependence of $A_k$, and some nuclear spins reach nearly full polarization while others remain minimally polarized.

It is worth noting that, aside from electron spin, hole~\cite{Eble2009Hole} or S-$T_{\pm}$ states~\cite{Johnson2005Triplet,Petta2008Dynamic,Yao2010Feedback} can also be utilized. Although the fluctuation of the Overhauser field (caused by nuclear spins) is not significantly suppressed, the electron polarization decay time can be slightly extended due to the protective effect of highly polarized nuclear spins. 
To illustrate, the averaged Overhauser field $\overline{h_o^z} = \langle h_o^z \rangle$ (in QDs, the averaged polarization ratio $p$ reaches $\sim$ 0.1-0.7~\cite{Bracker2005Optical,Lai2006Knight,Braun2005Direct,Tartakovskii2007Nuclear,Ono2004Nuclear,Wald1994Local}) is given by:
\begin{equation}\label{eq:s5}
\overline{h_o^z} = \frac{1}{2} \sum_k A_k p_k
\end{equation}
and its fluctuation $\Delta h_o^z$ is calculated in Eq.~\ref{eq:s3}. For GaAs QD with $10^6$ nuclear spins, $\overline{h_o^z}\sim 5$T and $\Delta h_o^z/\sqrt N \sim 5$mT when fully polarizing nuclear-spin bath~\cite{Koppens2005Control}. Note that in high polarization scenarios, dark states may exist to prevent further polarization~\cite{Imamo2003Optical,Taylor2003Controlling,Yu2013nuclear}.

\section{Bloch rotation and Fer expansion theory}\label{sup2}
For the hybrid dynamical decoupling (Hybrid DD) protocol, we apply Bloch rotation theory by introducing the Overhauser field $\overline{\mathbf{h}}_o=\sum_{i=1}^NA_k \langle\textbf{I}_k\rangle$. When the Overhauser field along $z$ axis is polarized using the NsBP method, we denote it as $\overline{\mathbf{h}_o^z}$.  Simultaneously, let $\mathbf{b}_e^z$ be the external bias magnetic field along the $z$-axis, and for convenience, define $a\equiv \gamma b_e^z+\overline{h_o^z}$, $b\equiv h_o^x$, $c\equiv h_o^y$. The effective bias field direction is given by $\mathbf{\hat{n}}=(n_x, n_y, n_z)$, where $n_z=\frac{a}{\sqrt{a^2+b^2+c^2}}$, $n_x=\frac{b}{\sqrt{a^2+b^2+c^2}}$, and $n_y=\frac{c}{\sqrt{a^2+b^2+c^2}}$.

In the Uni-DD protocol~\cite{Yao2019Uniaxial}, the dynamical operator $U_{\tau}$ provides the rotation angle $\theta=\tau\sqrt{a^2+b^2+c^2}$ during the pulse delay $\tau$. Hence, we have
\begin{eqnarray}
U_{2\tau}&=&Y(\pi)U_{\tau}Y(\pi)U_{\tau}\\
&=&e^{-i\theta S_{n,-}}e^{-i\theta S_{n,+}}.
\end{eqnarray}
Here, the decoupling pulses engineer the dynamic evolution direction for the electron spin in the Bloch sphere, and the engineered dynamic operators are $S_{n,-}\equiv -n_zS_z-n_xS_x+n_yS_y, and S_{n,+}\equiv n_zS_z+n_xS_x+n_yS_y$. We next expand it in a two-energy frame:
\begin{eqnarray}
\hspace*{-0.5cm} e^{-i\theta S_{n,-}}=\begin{pmatrix}
          \cos\frac{\theta}{2}+in_z\sin\frac{\theta}{2} & (in_x-n_y)\sin \frac{\theta}{2} \\  (in_x+n_y)\sin\frac{\theta}{2} & \cos\frac{\theta}{2}-in_z\sin\frac{\theta}{2}
           \end{pmatrix}
\end{eqnarray}
\begin{eqnarray}
\hspace*{-0.5cm} e^{-i\theta S_{n,+}}=\begin{pmatrix}
          \cos\frac{\theta}{2}-in_z\sin\frac{\theta}{2} & -(in_x+n_y)\sin\frac{\theta}{2} \\  (-in_x+n_y)\sin\frac{\theta}{2} & \cos\frac{\theta}{2}+in_z\sin\frac{\theta}{2}
           \end{pmatrix}
\end{eqnarray}
The magic condition can be divided into two cases: (1) when the bath is not polarized and the mean of $h_o^z$ is zero; and (2)when the bath is partially polarized and $\overline{h_o^z}$ takes a non-zero value. Next, we consider the second case; the first one simply assumes $\overline{h_o^z} = 0$ in a similar analysis below.
Since the ``magic" condition ensures $\theta \approx 2\pi$, and $n_z$ ($n_z \approx 1$) greatly exceeds $n_x \approx n_y \approx 0$, neglecting the second-order terms like $n_x n_y$, we obtain:
\begin{eqnarray}~\label{eq:s10}
 U_{2\tau}&=&
          \begin{pmatrix}
          1 & -n_y\theta^{'}\\ n_y\theta^{'} & 1
          \end{pmatrix}=\mathbb{I}_2-2in_y\theta^{'}S_y
 \end{eqnarray}
where $\theta^{'}$ is the small angle in the first-order Taylor expansion, and $\mathbb{I}_2$ is the $2 \times 2$ identity matrix. Hence, $U_{2n\tau} \approx \mathbb{I}_2$ guarantees long-time coherence in $n$-cycle QD-based quantum memory.
Note that the small term $2h_o^y / (b_e^z + \overline{b_o^z}) S_y$ of $U_{2\tau}$ provides better decoupling compared to the term $ 2h_o^y / b_e^z S_y$ in the Uni-DD protocol.

Similarly, using the Fer~\cite{Fer1958R,takegoshi2015comparison} and Magnus expansion methods, we can explain the hybrid DD protocol. With the same parameters as outlined above, the evolution operator for a short period $U_{2\tau}$ is given by $U_{2\tau} = U_1 U_0$ with the engineered Hamiltonian,
\begin{eqnarray}
H_0&=&\gamma b_e^zS_z+\overline{h_o^z}S_z+h_o^xS_x+h_o^yS_y,\\
H_1&=&-\gamma b_e^zS_z-\overline{h_o^z}S_z-h_o^xS_x+h_o^yS_y.
\end{eqnarray}
In the interaction picture, with the operator $R = \exp(-i t (b_e^z + \overline{h_o^z}) S_z) = \exp(-i a t S_z)$, we have:
$H_0^I=R^\dag H_0R=h_xS_x\cos(at)-h_xS_y\sin(at)+h_yS_y\cos(at)+h_yS_x\sin(at)$, and
$H_1^I=RH_1R^\dag=-h_xS_x\cos(at)-h_xS_y\sin(at)+h_yS_y\cos(at)-h_yS_x\sin(at)$,
hence,
 \begin{eqnarray}
  U_{2\tau}&=&RU_1^IR^\dag U_0^I.
 \end{eqnarray}
Under the magic condition $(\omega + \overline{h_o^z})\tau = 2\pi$, the Fer expansion can be applied to $U_{0}^I$ and $U_{1}^I$~\cite{Fer1958R, takegoshi2015comparison, Yao2019Uniaxial} using the following evolution operator $U_{\tau}$:
\begin{eqnarray}\label{eq:utauqd}
  U_{\tau} &=& R {\cal T}:\exp(-i\int_0^\tau H_R(t') dt') \nonumber \\
    &\approx & \exp(-i\tau H_{F,1})\exp(-i\tau H_{F,0}).
\end{eqnarray}
where $R=I$, and
\begin{widetext}
\begin{eqnarray}
H_{F,1} &=& \frac{S_x (\overline{h_o^z} h_o^x + h_o^x \overline{h_o^z}) + S_y (\overline{h_o^z} h_o^y + h_o^y \overline{h_o^z}) + S_z [(h_o^x)^2 + (h_o^y)^2]}{2\omega} 
        + i \frac{h_o^x h_o^y - h_o^y h_o^x}{4\omega}, \\
H_{F,0} &=& \overline{h_o^z} S_z \quad \text{for} \quad U_{0}^I,  \\
H_{F,1} &=& \frac{-S_x (\overline{h_o^z} h_x + h_x \overline{h_o^z}) + S_y (\overline{h_o^z} h_y + h_y \overline{h_o^z}) - S_z [(h_o^x)^2 + (h_o^y)^2]}{2\omega} 
        + i \frac{h_o^x h_o^y - h_o^y h_o^x}{4\omega}, \\
H_{F,0} &=& -\overline{h_o^z} S_z \quad \text{for} \quad U_{1}^I.
\end{eqnarray}
\end{widetext}
Finally, the average Hamiltonian theory can be exploited to expand $U_{2\tau}$~\cite{Haeberlen1968Coherent, haeberlen1976high, mehring1983principles, takegoshi2015comparison}:
\begin{widetext}
\begin{eqnarray}\label{eq:s19}
U_{2\tau} &=& \mathbb{I}_2 - 2i \tau \left [ \frac{S_y (\overline{h_o^z} h_o^y + h_o^y \overline{h_o^z})}{2(\omega + \overline{h_o^z})} + i \frac{(h_o^x h_o^y - h_o^y h_o^x)}{4(\omega + \overline{h_o^z})} \right ],
\end{eqnarray}
\end{widetext}
where $\overline{H}=S_y(\overline{h_o^z}h_o^y)/(\omega + \overline{h_o^z})+i(h_xh_o^y-h_o^yh_o^x)/[4(\omega+ \overline{h_o^z})]$ and $\overline{h_o^z}=\gamma b_o^z$. Note that the first two terms of Eq.~\ref{eq:s19} keeps similarly with Eq.~\ref{eq:s10} in mean-field level. Compared to the free Hamiltonian of Eq.~\ref{eq:s1} in the QD system, where $\overline{H} = h_o^x S_x + h_o^y S_y + h_o^z S_z$, the effective coupling constant in the average Hamiltonian is reduced by a small factor $|h_o^z|/(\omega + \overline{h_o^z})$, rather than $|h_o^z|/\omega$ in the Uni-DD method. At the same time, three coupling directions are reduced to one direction along the $y$-axis in both the hybrid DD and the Uni-DD methods. The last term introduces the intra-bath dynamics, which can affect the electron spin dynamics over long timescales.

\section{the proposal for electronic spin quantum memory in the QD}\label{sup3}
\begin{figure*}[!htpb]
\centering
\includegraphics[width=3.4in]{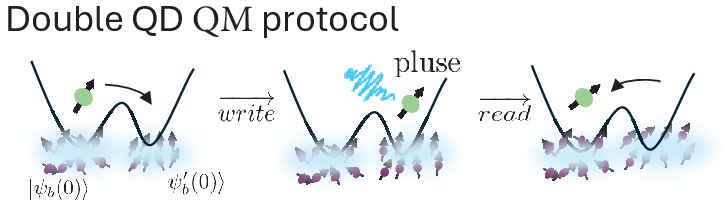}
\caption{\label{fig:sa} Electronic spin quantum memory sketch: The process begins with an electron being injected into  the right QD, where the nuclear-spin bath is partially polarized, while the bath in the left QD remains unchanged with $| \psi_b(0)\rangle$. This results in the collective bath state $|\psi_b^{'}(0)\rangle$ in the right QD. Then, electronic spin is dynamically preserved by pulse sequences, including hybrid DD (Uni-DD) and its symmetrized hybrid DD (Uni-DD) and concatenated hybrid DD (Uni-DD). Finally, the electron spin is ejected into the left QD to read its memory information.}
\end{figure*}

In Fig.~\ref{fig:sa}, we give an illustrated quantum memory based on electronic spin in double QDs. The distance between left QD and right QD is $200\sim500 nm$~\cite{Petta2008Dynamic}, which ensures that the electronic spin dynamics in the left and right quantum dots are well isolated from each other, thereby allowing each dot to function as an independent subsystem. 
To initialize the quantum memory, we employ repeated DNP process to prepare the nuclear-spin bath in the right quantum dot into a desired initial state, denoted as $|\psi_b^{'}(0)\rangle$. After this initialization, the quantum memory operation proceeds as follows: first, the electron is transferred from the left dot to the right dot (the write process); second, the quantum information encoded in the central electronic spin is preserved by applying dynamical decoupling (Uni-DD) pulse sequences during the storage period; finally, the electron spin is transferred back from the right dot to the left dot (the read process).

Specifically, we investigate three types of uniaxial DD (Uni-DD) protocols: (1) Uni-DD, which consists of two instantaneous $\pi_y$ pulses applied along the $y$-axis within each unit cycle.
(2) Symmetrized Uni-DD (SUni-DD) – defined as $[\overline{U}\tau Y \overline{U}\tau U_\tau Y U_\tau]$, where $U_\tau$ represents the unitary evolution under a static bias magnetic field, and $\overline{U}_\tau$ denotes the reversed evolution with an opposite bias field ($\omega = -\gamma B$). This symmetrization improves the robustness of the protocol against systematic errors.
(3) Concatenated Uni-DD (CUni-DD) – constructed hierarchically from SUni-DD. At the second concatenation level, denoted $C_2$, the sequence is $[\overline{C}_1 Y \overline{C}_1 C_1 Y C_1]$, where $C_1 =$ SUni-DD. Each cycle of CUni-DD involves 16 $\pi_y$ pulses.
In the absence of nuclear spin polarization, the decoupling performance of these sequences has been systematically analyzed in the appendix of Ref.~\cite{Yao2019Uniaxial}. However, when the nuclear spins are dynamically polarized via DNP, the performance of our proposed hybrid DD protocol—which combines Uni-DD with bath polarization—exceeds that of the standard Uni-DD family. This advantage arises because nuclear polarization suppresses bath fluctuations, thereby relaxing the requirements on the external magnetic field.

Notably, the symmetrized and concatenated versions of the hybrid DD protocol preserve the same mathematical forms as SUni-DD and $C_2$, respectively. The crucial difference between the hybrid DD and Uni-DD is that the external bias field required for satisfying the modified “magical condition” is significantly reduced in the presence of bath-spin polarization. The actual value of the field can be estimated by considering both the Overhauser field polarization ratio and the magic condition.

\end{appendix}

%
\end{document}